\documentclass[12pt]{iopart}
\usepackage{graphicx}
\usepackage{dcolumn}
\usepackage{amssymb}
\usepackage{bm}

\begin{document}

\title{Spontaneous symmetry breaking in Loop Quantum Gravity}

\author{G Helesfai}

\address{\dag\ Institute for Theoretical Physics, E\"otv\"os University,P\'azm\'any P\'eter s\'et\'any 1/A, H-1117 Budapest, Hungary}
\eads{\mailto{heles@manna.elte.hu}}

\begin{abstract}

In this paper we investigate the question how spontaneous symmetry breaking works in the framework of 
Loop Quantum Gravity and we compare it to the results obtained in the case of the Proca field, where 
we were able to quantize the theory in Loop Quantum Gravity without introducing a Higgs field. We 
obtained that the Hamiltonian of the two systems are very similar, the only difference is an extra 
scalar field in the case of spontaneous symmetry breaking. This field can be identified as the field 
that carries the mass of the vector field. In the quantum regime this becomes a well defined operator, 
which turns out to be a self adjoint operator with continuous spectrum. To calculate the spectrum we 
used a new representation in the case of the scalar fields, which in addition enabled us to rewrite 
the constraint equations to a finite system of linear partial differential equations. This made it 
possible to solve part of the constraints explicitly. 

\end{abstract}


\maketitle

\section{Introduction}

Currently spontaneous symmetry breaking is the most accepted tool to define mass to particles. 
Its success can be observed especially in the case of vector fields since their original Lagrangian - 
the Proca Lagrangian - is non-renormalizable. In \cite{Proca} we showed how one can quantize the 
massive vector field in Loop Quantum Gravity without spontaneous symmetry breaking. The main problem 
was that the Proca field had a second class constraint algebra which made it almost impossible to 
apply the framework of LQG. But with the help of symplectycal embedding one could eliminate these 
difficulties. Now the question arises what is the difference between the two theories. To study this 
one first has to apply the framework of LQG to a system where spontaneous symmetry breaking is used to 
generate mass for a U(1) vector field. This is done in sections 2 - classical theory and 3+1 
decomposition - and 3 - regularisation and quantization. In section 4 we introduce a new basis 
for the scalar fields which is motivated by the fact that these are eigenstates of the configuration 
variables. It turns out that with the help of this new we are able to (partially) solve the constraints 
of the theory - this is done in section 5. Here we also analyze special solutions in order to understand 
the role of the scalar field. In particular we find that some of these are almost identical to the 
solutions obtained for the Proca field, thus we are able to relate the two theories. Section 6 deals 
with the ``mass operator'' and its properties, concentrating especially on those cases where the 
eigenvalues of this operator can be identified with the mass parameter of the Proca field. We obtain that 
in a sense the case of the symmetry breaking is a linear combination of infinite Proca theories. 

\section{Classical theory}

In this section we will analyze the general framework of spontaneous symmetry breaking from a Hamiltonian 
perspective. In the first subsection we derive the 3+1 decomposition of the theory, while in the second 
we first have the U(1) field a VEV then perform the 3+1 decomposition (this is useful because the similarities 
between the Proca field and spontaneous symmetry breaking become more transparent).

\subsection{Symmetric theory}

For simplicity we use the Lagrangian of a U(1) vector field (electromagnetic field) coupled to gravity and a 
U(1) complex scalar field on a space-time manifold M. The Lagrangian of the matter part is
\begin{eqnarray}\hspace{-2cm}L^{mat}=\int_M d^4x\sqrt{-g}\left(-\frac{1}{4}\underline{F}^{(4)}_{ab}\underline{F}^{(4)ab}-\frac{1}{2}D^{(4)}_{a}\Phi^* D^{(4)a}\Phi-\frac{1}{4}\mu(\Phi^* \Phi-a^2)^2\right),
\end{eqnarray}
where
\begin{eqnarray}
\Phi &=& \Re\Phi+i\Im\Phi\nonumber\\D^{(4)}_{a}\Phi &=& (\partial^{(4)}_a+ie\underline{A}^{(4)}_a)\Phi\nonumber
\end{eqnarray}
and $\mu$ and $a$ are positive constants. To distinguish between the electromagnetic field and the gravitational 
field the variables of the former are underlined.\\
First let us make the 3+1 decomposition. Introduce on the space-time manifold $M$ a smooth function $t$ whose 
gradient is nowhere vanishing and a vector field $t^a$ with affine parameter $t$ satisfying $t^a\nabla_a t=1$. 
This gives a foliation of space-time, i. e. each $t$ defines a 3-dimensional hypersurfice $\Sigma_t$. Let us 
decompose $t^a$ into its normal and tangential part
\begin{eqnarray}
t_a=Nn_a+N_a,
\end{eqnarray}
where $n_a$ is the unit normal of the hypersurfice $\Sigma_t$, $N$ is the lapse function, $N_a$ is the shift vector. 
Define the induced, positive-definite metric on $\Sigma_t$ via
\begin{eqnarray}
q_{ab}=g_{ab}+n_a n_b\label{3and1}
\end{eqnarray}
As it was done in \cite{newvar} and \cite{Proca} we define the pull-backs 
$\underline{A}_a=q^b_a \underline{A}^4_b\ ,\ D_a=q^b_a D^4_b$ and define 
$\underline{A}_0=t^a \underline{A}^4_a, A_0^i=t^a A^{i4}_a$. Substituting these into the Lagrangian one obtains
\begin{eqnarray}
\hspace{-2cm}L^{mat}=\int dt\int d^3x \left(\frac{N}{\sqrt{q}}q_{ab}\frac{\underline{E}^a \underline{E}^b-\underline{B}^a \underline{B}^b}{2}\right)-\frac{1}{4}N\sqrt{q}\mu(\Phi^* \Phi-a^2)^2-\nonumber\\
\hspace{-2cm}-\frac{1}{2}N\sqrt{q}\left[q^{cd}D_c\Phi^* D_d\Phi-\frac{(\mathcal{L}_t\Phi^*-ieA^0\Phi^*-N^a(D_a\Phi)^*)(\mathcal{L}_t\Phi+ieA^0\Phi-N^b D_b\Phi)}{N^2}\right],\nonumber\\
\end{eqnarray}
where
\begin{eqnarray}
\underline{E}_a=\frac{\sqrt{q}}{N}(\mathcal{L}_t \underline{A}_c-D_c \underline{A}^0-\epsilon_{abc} \underline{B}^b N^c)
\end{eqnarray}
is the electric field and $\underline{B}_a$ is the magnetic field. We now define the canonical momenta
\begin{eqnarray}
\underline{\Pi}^a &=& \frac{\delta L}{\delta \mathcal{L}_t \underline{A}_a}=\underline{E}^a\\
\pi &=& \frac{\delta L}{\delta \mathcal{L}_t \Phi}=\sqrt{q}\frac{\mathcal{L}_t\Phi^*-ie\underline{A}^0\Phi^*-N^a(D_a\Phi)^*}{N}\\
\pi^* &=& \frac{\delta L}{\delta \mathcal{L}_t \Phi^*}=\sqrt{q}\frac{\mathcal{L}_t\Phi+ie\underline{A}^0\Phi-N^b D_b\Phi}{N}.
\end{eqnarray}
Finally we perform the Legendre transformation to arrive to the Hamiltonian:
\begin{eqnarray}
\hspace{-2cm}H^{mat} &=& \int d^3 x (N\mathcal{H}^{mat}+N^a\mathcal{H}^{mat}_a+A_0 \underline{G})\label{ham}\\
\hspace{-2cm}\mathcal{H}^{mat} &=& q_{ab}\left(\frac{\underline{E}^a \underline{E}^b+\underline{B}^a \underline{B}^b}{2\sqrt{q}}\right)+\frac{\pi\pi^*}{2\sqrt{q}}+\frac{1}{2}\sqrt{q}(\mathcal{D}_a\Phi)^* \mathcal{D}_a\Phi+\frac{1}{4}\sqrt{q}\mu(\Phi^* \Phi-a^2)^2\\
\hspace{-2cm}\mathcal{H}^{mat}_a &=& \epsilon_{abc} \underline{B}^c \underline{E}^b+\pi\mathcal{D}_a\Phi+\pi^* (\mathcal{D}_a\Phi)^*\\
\hspace{-2cm}\underline{G} &=& D_a \underline{E}_a+ie(\pi^* \Phi^* -\pi\Phi)\nonumber
\end{eqnarray}
where $\mathcal{H}^{mat},\mathcal{H}^{mat}_a$ and $\underline{G}$ are the matter contributions to the 
Hamiltonian- and diffeomorphism (modulo gauge transformations) constraints, and the electromagnetic 
Gauss constraint.\\
The (non-smeared, non-trivial) Poisson-brackets are:
\begin{eqnarray}
\{\pi,\Phi\} &=& \delta(x,y)\\
\{\pi^*,\Phi^*\} &=& \delta(x,y)\\
\{E_a,A^b\} &=& \delta_a^b\delta(x,y)
\end{eqnarray}
Since we are going to do symmetry breaking with the help of the scalar field, we write here  
the transformation rules for the scalar fields and their canonical momenta with respect to infinitesimal 
gauge transformation:
\begin{eqnarray}
\{\underline{G}(\Lambda),\Phi\} &=& -ie\Lambda\Phi\nonumber\\
\{\underline{G}(\Lambda),\Phi^*\} &=& ie\Lambda\Phi^*\nonumber\\
\{\underline{G}(\Lambda),\pi\} &=& ie\Lambda\pi\nonumber\\
\{\underline{G}(\Lambda),\pi^*\} &=& -ie\Lambda\pi^*\nonumber
\end{eqnarray}

Before we continue, there are a few interesting observations that should be mentioned here:

\begin{enumerate}

\item[-] All the constraints are real and only the scalar fields and their canonical momenta are 
represented by complex variables (note that in the Hamiltonian picture $\Phi$ and $\Phi^*$ are independent variables).
\item[-] The transformation $\Phi\leftrightarrow\Phi^*,\ \pi\leftrightarrow\pi^*$ is a canonical transformation.
\item[-] The true diffeomorphism constraint $\mathcal{H}^{mat}_a+\underline{A}_a\underline{G}$ is 
independent of the coupling constant e (it contains partial derivatives only).
\item[-] This system has a first class constraint algebra, further more all the components of 
$\mathcal{H}_{mat}$ are gauge invariant respectively.

\end{enumerate}

\subsection{New variables}

In spontaneous symmetry breaking first we introduce new fields $\eta$ and $\Theta$ in the following way:
\begin{eqnarray}
\Phi(x):=(a+\eta(x))\exp\left(i\frac{\Theta(x)}{a}\right).
\end{eqnarray}

These variables are useful because the U(1) symmetry of the theory becomes more transparent. If we substitute 
this into the Lagrangian, we obtain

\begin{eqnarray}
L_{mat}=\int d^4 x \sqrt{-g}\left[-\frac{1}{4}\underline{F}^{(4)}_{ab}\underline{F}^{(4)ab}-\frac{1}{2}\partial^{(4)}_a \eta\partial^{4)a}\eta-\right.\nonumber\\
\left.-\frac{1}{2}(a+\eta)^2\left(\frac{\partial^{(4)}_a \Theta}{a}+e\underline{A}_a\right)\left(\frac{\partial^{(4)}_a\Theta}{a}+e\underline{A}_a\right)-\frac{1}{4}\mu\eta^2(2a+\eta)^2\right].
\end{eqnarray}

If we compare this with the action of the symplectically embedded Proca field we immediately 
recognize the similarities between the two theories. The main difference is that where the Proca theory 
had a parameter (m), now we have a field ($\eta+a$). We wish to see how the Hamiltonian looks like in 
terms of the new variables. To do this, first we do the 3+1 decomposition of the above Lagrangian. 
Repeating the steps of the previous section first we define the canonical momenta

\begin{eqnarray}
\underline{\Pi}^a &=& \frac{\delta L}{\delta \mathcal{L}_t \underline{A}_a}=\underline{E}^a\\
\pi_\eta &=& \frac{\delta L}{\delta \mathcal{L}_t \eta}=\sqrt{q}\frac{\mathcal{L}_t\eta-N^a\partial_a\eta}{N}\\
\pi_\Theta &=& \frac{\delta L}{\delta \mathcal{L}_t \Theta}=\left(\frac{a+\eta}{a}\right)^2\sqrt{q}\left(\frac{\mathcal{L}_t\Theta-N^a\partial_a\Theta+aeA_0-aeN^a \underline{A}_a}{N}\right)
\end{eqnarray}

and after the Legendre-transformation we obtain the Hamiltonian

\begin{eqnarray}
\hspace{-2cm}H_{mat} &=& \int d^3 x (N\mathcal{H}_{mat}+N^a\mathcal{H}^{mat}_a+A_0 \underline{G})\label{hamnew}\\
\hspace{-2cm}\mathcal{H}_{mat} &=& q_{ab}\left(\frac{\underline{E}^a \underline{E}^b+\underline{B}^a \underline{B}^b}{2\sqrt{q}}\right)+\frac{\pi_\eta^2}{2\sqrt{q}}+\frac{1}{2}\sqrt{q}\partial_a\eta\partial_a\eta+\nonumber\\
\hspace{-2cm}&+&\left(\frac{a}{a+\eta}\right)^2\frac{\pi_\Theta^2}{2\sqrt{q}}+\frac{1}{2}\sqrt{q}\left(\frac{a+\eta}{a}\right)^2(\partial_a\Theta+ae\underline{A}_a)^2+\frac{1}{4}\sqrt{q}\mu\eta^2(2a+\eta)^2\\
\hspace{-2cm}\mathcal{H}^{mat}_a &=& \epsilon_{abc} \underline{B}^c \underline{E}^b+\pi_\eta\partial_a\eta+\pi_\Theta\partial_a\Theta+ae\underline{A}_a\pi_\Theta\\
\hspace{-2cm}\underline{G} &=& D_a \underline{E}_a-ae\pi_\Theta
\end{eqnarray}

The (non-smeared, non-trivial) Poisson-brackets:

\begin{eqnarray}
\{\pi_\eta,\eta\} &=& \frac{1}{2}\delta(x,y)\\
\{\pi_\Theta,\Theta\} &=& \frac{1}{2}\delta(x,y)\\
\{\underline{E}_a,\underline{A}^b\} &=& \delta_a^b\delta(x,y)
\end{eqnarray}

Now if one compares the Hamiltonian \eref{hamnew} with the original \eref{ham}, it is easy to see that the 
two are connected with the help of the following canonical transformation:

\begin{eqnarray}
\Phi &:=& (a+\eta)\exp\left(\frac{i\Theta}{a}\right)\nonumber\\
\Phi^* &:=& (a+\eta)\exp\left(-\frac{i\Theta}{a}\right)\nonumber\\
\pi &:=& \left(\pi_\eta-\frac{ia}{a+\eta}\pi_\Theta\right)\exp\left(-\frac{i\Theta}{a}\right)\nonumber\\
\pi^* &:=& \left(\pi_\eta+\frac{ia}{a+\eta}\pi_\Theta\right)\exp\left(\frac{i\Theta}{a}\right)\nonumber
\end{eqnarray}

Further more the above system is very similar to the case of the symplectically embedded Proca-field. 
To see this, let us introduce the canonical transformation $\pi_\Theta\ \to\ \frac{\pi_\Theta}{ea},\ \Theta\ \to\  ea\Theta$ 
and define $m^2=e^2(a+\eta)^2$. Then we will obtain exactly the Hamiltonian of \cite{Proca}, with the 
exception of a potential term. There are two major differences: there is an extra dynamical scalar field 
in the theory and the ``mass'' is constructed from the field $\eta$. The later will be quite important 
since after quantization all the fields will become operators so we can define a ``mass operator'', which 
spectrum can be identified as the mass spectrum (in \cite{Proca} the mass was a parameter of the theory).   

\subsection{Classical symmetry breaking}

In quantum field theory we use the unitary gauge to do gauge fixing. In the U(1) case this means we introduce 
the gauge-fixed vector field 

\begin{eqnarray}
\underline{\tilde{A}}^{(4)}_a(x):=\underline{A}^{(4)}_a(x)-\frac{1}{ea}\partial^{(4)}_a\Theta(x).
\end{eqnarray}

Substituting this into the Lagrangian we get

\begin{eqnarray}
\tilde{L}_{mat}=\int d^4 x \sqrt{-g}\left[-\frac{1}{4}\underline{\tilde{F}}^{(4)}_{ab}\underline{\tilde{F}}^{(4)ab}-\frac{1}{2}\partial^{(4)}_a\eta\partial^{(4)a}\eta-\frac{1}{2}e^2(a+\eta)^2\underline{\tilde{A}}^{(4)}_a \underline{\tilde{A}}^{(4)a}-\right.\nonumber\\
\left.-\frac{1}{4}\mu\eta^2(2a+\eta)^2\right]
\end{eqnarray}

Again we want to see how the Hamiltonian changes, so we do the 3+1 decomposition as we did in the previous sections. 
The canonical momenta will be 

\begin{eqnarray}
\underline{\tilde{\Pi}}^a &=& \frac{\delta L}{\delta \mathcal{L}_t \underline{\tilde{A}}_a}=\underline{\tilde{E}}^a\nonumber\\
\pi_\eta &=& \frac{\delta L}{\delta \mathcal{L}_t \eta}=\sqrt{q}\frac{\mathcal{L}_t\eta-N^a\partial_a\eta}{N},\nonumber
\end{eqnarray}

and the constraints will be

\begin{eqnarray}
\mathcal{H}_{mat} &=& q_{ab}\frac{\underline{\tilde{E}}^a \underline{\tilde{E}}^b+\underline{\tilde{B}}^a \underline{\tilde{B}}^b}{2\sqrt{q}}+\frac{\pi_r^2}{2\sqrt{q}}+\frac{1}{2}\sqrt{q}\partial_a\eta\partial_a\eta+\nonumber\\&+& \frac{1}{4}\sqrt{q}\mu\eta^2(2a+\eta)^2+\frac{1}{2}e^2\sqrt{q}(a+\eta)^2(\underline{\tilde{A}}_a\underline{\tilde{A}}^a+(\frac{\tilde{A}_0-N^a\underline{\tilde{A}}_a}{N})^2)\\
H^{mat}_a &=& \epsilon_{abc} \underline{\tilde{B}}^c \underline{\tilde{E}}^b+\pi_r\partial_a\eta+e^2\underline{\tilde{A}}_a\sqrt{q}(a+\eta)^2\frac{\tilde{A}_0-N^a\underline{\tilde{A}}_a}{N}\\
\underline{G} &=& D_a \underline{\tilde{E}}_a-e^2\sqrt{q}(a+\eta)^2\frac{\tilde{A}_0-N^a\underline{\tilde{A}}_a}{N}
\end{eqnarray}

\newpage

The (non-smeared, non-trivial) Poisson-brackets remain the same:

\begin{eqnarray}
\{\pi_r,\eta\} &=& \frac{1}{2}\delta(x,y)\\
\{\tilde{E}_a,\tilde{A}^b\} &=& \delta_a^b\delta(x,y)
\end{eqnarray}

If we compare this gauge fixed Hamiltonian with \eref{hamnew}, we can see that in the Hamiltonian 
formalism the gauge fixing is equivalent to the introduction of the following two constraints

\begin{eqnarray}
C_a &:=& \partial_a Arg(\Phi)=0\\
C &:=& a\pi_f-e\sqrt{q}(a+\eta)^2\frac{\tilde{A}_0-N^a\tilde{A}_a}{N}=0,
\end{eqnarray}

(the second is equivalent to $\mathcal{L}_t\Theta=0$). This is precisely the gauge we used in the case of 
the symplectically embedded Proca field. There we showed that in LQG gauge fixing is not necessary, in fact 
it makes the quantization extremely difficult if not impossible. So we will not fix the gauge, instead we 
will try to solve the constraint related to it.\\ \\
To conclude we summarize the most important results of this section.\\ \\
We checked how one can implement spontaneous symmetry breaking in the Hamiltonian formalism. It turned 
out that introducing new variables means a canonical transformation, while gauge fixing (as it was shown 
earlier e.g. in \cite{gfixing}) can be done by introducing new constraints. Interestingly these are exactly 
the same conditions which were introduced in the case of the Proca field in \cite{Proca}. Further more the 
Hamiltonian \eref{hamnew} is very similar to the symplectically embedded Proca Hamiltonian (\cite{Proca}, page 5), 
the only two exception is that we have an extra scalar field and the mass is not a parameter but defined 
with the help of the field $\eta$. 

\section{Quantization}

\subsection{Gauge fields}

Quantization of the gravitational and electromagnetic field can be treated on the same footing since 
both are gauge fields - the gauge group of the former is, in the Ashtekar variables (\cite{newvar}), SU(2) 
while the latter is a U(1) field. The detailed analysis of the method can be found in \cite{qsd}-\cite{quantdiff}, 
here we just sketch the main idea and the notations.\\
Let us consider a Yang-Mills gauge field with a compact gauge group G. The Hilbert-space can be constructed 
in the following way: let $\gamma$ be an oriented graph in $\Sigma$ with $e_1,\dots,e_E$ edges and $v_1,\dots,v_V$ 
vertices. Let $h_{e_i}$ be the holonomy of the G-valued connection of the field evaluated along the $e_i$ edge. 
Let us define a cylindrical function with respect to a $\gamma$ graph in the following way:

\begin{eqnarray}
f_{\gamma}(A):=f(h_{e_1},\dots,h_{e_E})
\end{eqnarray}

where $f_{\gamma}$ is a complex valued function mapping from $G^E$. The Hilbert-space of the Yang-Mills field is 
defined as the set of all cylindrical functions which are square-integrable with respect to a suitable measure 
(the Ashtekar-Lewandowski measure): 
\begin{eqnarray}
\mathcal{H}:=L_2(\bar{\mathcal{A}},d\mu_{AL,G})
\end{eqnarray}

In our case, $G=SU(2)\times U(1)$, so 

\begin{eqnarray}
\mathcal{H}_{G,YM}:=L_2(\bar{\mathcal{A}}_{SU(2)},d\mu_{SU(2)})\otimes L_2(\bar{\underline{\mathcal{A}}}_{U(1)},d\mu_{U(1)})\label{HilbertG}
\end{eqnarray}

In order to analyze the action of the Hamiltonian and to compute its kernel,it is convenient to introduce a 
complete orthonormal basis on the Hilbert-space\eref{HilbertG}.\\
On the space of $L_2(\bar{\mathcal{A}}_{SU(2)},d\mu_{SU(2)})$ these are called \textit{spin network functions} 
and defined as follows: let $\gamma\in\Sigma$ be a graph and denote its edges and vertices respectively by 
$(e_1,\dots,\e_N)$ and $(v_1,\dots,v_V)$. Associate a coloring to each edge defined by a set of irreducible 
representations $(j_1,\dots,j_N)$ of SU(2) (half-integers) and contractors $(\rho_1,\dots,\rho_V)$ to the vertices 
where $\rho_l$ is an intertwiner which maps from the tensor product of representations of the incoming edges at 
the vertex $v_l$ to the tensor product of representations of the outgoing edges. A spin network state is then 
defined as

\begin{eqnarray}
|T(A)\rangle_{\gamma,\vec{j},\vec{\rho}}:=\bigotimes_{i=1}^N j_i(h_{e_i}(A))\cdot\bigotimes_{k=1}^V\rho_k\label{spin1}
\end{eqnarray}

where $\cdot$ stands for contracting at each vertex $v_k$ the upper indices of the matrices corresponding to all 
the incoming edges and the lower indices of the matrices assigned to the outgoing edges with all the indices of 
$\rho_k$.\\
In the case of $L_2(\bar{\mathcal{A}}_{U(1)},d\mu_{U(1)})$ one must simply replace SU(2) with U(1) in the above 
definition - these are called \textit{flux network functions} (\cite{mq3},\cite{mq2}). Since U(1) is a commutative 
group, we will have the following definition:for each edge $e_i$ of the graph associate an integer $l_i$. Then the 
flux network function is defined as 

\begin{eqnarray}
|F(\underline{A})\rangle_{\gamma,\vec{l}}:=\prod_{i=1}^N (h_{e_i}(\underline{A}))^{l_i}
\end{eqnarray}   

What remains is to define the operators corresponding to the connection and the electric field on the Hilbert-space. 
If we want to implement the Poisson-brackets in the quantum theory in a diffeomorphism covariant way, we have to use 
smeared versions of these fields. In the case of gauge fields the natural candidates are the holonomy and the electric 
flux respectively:

\begin{eqnarray}
h_e(A)=\mathcal{P}\exp{\int_e A}\label{sm1}\\E(S)=\int_S*E\label{sm2},
\end{eqnarray}

where $e$ is a path and $S$ is a surface in $\Sigma$, and $*E$ is the dual of the electric field. Then the action 
of the corresponding operators will be defined as:

\begin{eqnarray}
\hat{h}_e(A)f(A):=h_e(A)f\\\hat{E}(S) f(A):=i\bar{h}\{E(S),f(A)\}\label{der}
\end{eqnarray}  

\subsection{Scalar field}

The crucial point of quantizing the scalar field is (see \cite{qsd5} or \cite{scalar1},\cite{scalar2}) that the 
field should be real valued. In our case the original variables are complex, but this does not cause significant 
difficulties since we can introduce new fields which are real, thus the usual techniques can be applied on them. 
The only non-trivial problem is an additional ambiguity which arises because this can be done more than one way. 
What we are going to do is introduce two kinds of different choices for the configuration variables and the momentum 
operators.\\ \\
{\bf Case A:} 

The most natural choice is to define the operators with the help of the real and imaginary parts of the fields. 
Let v be a vertex of a graph $\gamma$ with coordinates $x_v$. Then let

\begin{eqnarray}
U(\lambda,v) &:=& \exp(i\lambda\Re(\Phi(x_v)))\\
\bar{U}(\delta,v) &:=& \exp(i\delta\Im(\Phi(x_v))),
\end{eqnarray}

where $\lambda$ and $\delta$ are arbitrary real numbers which are required because otherwise the quantization 
would not be general enough (see \cite{scalar1} and \cite{scalar2} for details). The variables for the momentum 
operator should be (B is an open ball in $\Sigma$)

\begin{eqnarray}
\Pi(B) &=& \int_B d^3 x \Re(\pi)\nonumber\\
\bar{\Pi}(B) &=& \int_B d^3 x \Im(\pi)\nonumber
\end{eqnarray}

and thus the Poisson-brackets of the variables will be

\begin{eqnarray}
\{\Pi(B),U(\lambda,v)\} &=& \delta_{v\cap B,v}\frac{i\lambda}{2}U(\lambda,v)\\
\{\bar{\Pi}(B),\bar{U}(\delta,v)\} &=& -\delta_{v\cap B,v}\frac{i\delta}{2}\bar{U}(\delta,v)\\
\{\Pi(B),\bar{U}(\delta,v)\} &=& \{\bar{\Pi}(B),U(\lambda,v)\}=0\label{theycommute}
\end{eqnarray}

The transformation rules of these quantities with respect to the (smeared) gauge transformation 
($\underline{G}(\Lambda)=\int d^3x\underline{G}\Lambda $) are:

\begin{eqnarray}
\{\underline{G}(\Lambda),U(\lambda,v)\} &=& ie\Lambda\lambda\Im(\Phi(x_v))U(\lambda,v)\nonumber\\
\{\underline{G}(\Lambda),\bar{U}(\delta,v)\} &=& -ie\Lambda\delta\Re(\Phi(x_v))\bar{U}(\delta,v)\nonumber
\end{eqnarray}

Since we have two fields, the Hilbert space for the scalar field is a tensor product 
$\mathcal{H}_{sc}=\mathcal{H}(U)\bigotimes\mathcal{H}(\bar{U})$, where the Hilbert spaces 
$\mathcal{H}(U)$ and $\mathcal{H}(\bar{U})$ are the linear combination of the following monomonials: 
let $\underline{v}=v_1,\dots,v_N$ be the set of vertices for some $\gamma$ graph and let $\underline{\lambda}$ and 
$\underline{\bar{\delta}}$ be two sets of real numbers, each pair associated to a vertex. Then a basic element 
of $\mathcal{H}(U)$ is constructed as follows:

\begin{eqnarray}
|\underline{\lambda}\rangle_\gamma=\prod_{k=1}^N U(\lambda_k,v_k)\label{phaseH1}
\end{eqnarray}

In a similar fashion

\begin{eqnarray}
|\bar{\underline{\delta}}\rangle_\gamma=\prod_{k=1}^N \bar{U}(\bar{\delta}_k,v_k)\label{phaseH2}
\end{eqnarray}

will be a basic element is $\mathcal{H}(\bar{U})$. Both $|\underline{\lambda}\rangle_\gamma$ and 
$|\bar{\underline{\delta}}\rangle_\gamma$ form a complete orthonormal basis, that is 
$_{\gamma'}\langle\underline{\lambda}'|\underline{\lambda}\rangle_\gamma=\delta_{\underline{\lambda},\underline{\lambda}'}\delta_{\gamma,\gamma'}$ and the same is true for $|\bar{\underline{\delta}}\rangle_\gamma$.\\
Thus elements of $\mathcal{H}_{sc}$ are linear combinations of of monomonials 
$|\underline{\lambda}\rangle_\gamma |\bar{\underline{\delta}}\rangle_\gamma$.\\
The operators are defined in the same way as in the case of gauge fields:

\begin{eqnarray}
\hat{U}(\lambda,v)|\underline{\lambda}\rangle_\gamma &:=& U(\lambda,v)|\underline{\lambda}\rangle_\gamma\\
\hat{\Pi}(B)|\underline{\lambda}\rangle_\gamma &:=& i\bar{h}\{\Pi(B),|\underline{\lambda}\rangle_\gamma\}\\
\hat{\bar{U}}(\delta,v)|\underline{\bar{\delta}}\rangle_\gamma &:=& \bar{U}(\delta,v)|\underline{\bar{\delta}}\rangle_\gamma\\
\hat{\bar{\Pi}}(B)|\underline{\bar{\delta}}\rangle_\gamma &:=& i\bar{h}\{\bar{\Pi}(B),|\underline{\bar{\delta}}\rangle_\gamma\}
\end{eqnarray}

Because the U(1) group is commutative, the action of the operators are very simple:

\begin{eqnarray}
\hat{\Pi}(B)|\underline{\lambda}\rangle_\gamma=-\frac{\bar{h}}{2}\sum_{v_j\in B}\lambda_j|\underline{\lambda}\rangle_\gamma\\
\hat{U}(\lambda,v)|\underline{\lambda}\rangle_\gamma=|\underline{\lambda'}\rangle_\gamma\nonumber\\
\lambda'_i=\lambda_i+\delta_{v,v_i}\lambda
\end{eqnarray}

and similar expressions hold for $\hat{\bar{U}}(\delta,v)$ and $\hat{\bar{\Pi}}(B)$. Also, because of \eref{theycommute} 
$\hat{\bar{\Pi}}(B)|\underline{\lambda}\rangle_\gamma=\hat{\Pi}(B)|\underline{\bar{\delta}}\rangle_\gamma=0$.\\ \\
{\bf Case B:} 

Another way is to use the absolute value and the argument of $\Phi$. Actually these are equal (up to constant factors) 
with the fields $\eta$ and $\Theta$ respectively, so we suggest the following operators for the multiplication 
operators:

\begin{eqnarray}
U_\eta(\lambda,v) &:=& \exp(i\lambda\eta(x_v))\\
U_\Theta(\delta,v) &:=& \exp(i\delta\frac{\Theta(x_v)}{a}).
\end{eqnarray}

For the momentum operators it is plausible to use the quantities $\pi_\eta$ and $\pi_\Theta$ instead of $\Re(\Pi)$ 
and $\Im(\Pi)$:
\begin{eqnarray}
\Pi_\eta(B) &=& \int_B d^3x \pi_\eta\nonumber\\
\Pi_\Theta(B) &=& a\int_B d^3x \pi_\Theta\nonumber
\end{eqnarray}

The Poisson-brackets of these variables are a bit different then in case A:

\begin{eqnarray}
\{\Pi_\eta(B),U_\eta(\lambda,v)\} &=& i\frac{1}{2}\lambda\delta_{v\cap B,v} U_\eta(\lambda,v)\\
\{\Pi_\Theta(B),U_\Theta(\delta,v)\} &=& i\frac{1}{2}\delta\delta_{v\cap B,v} U_\Theta(\delta,v)\\
\{\Pi_\eta(B),U_\Theta(\delta,v)\} &=& \{\Pi_\Theta(B),U_\eta(\lambda,v)\}=0
\end{eqnarray}

The transformation rule for these variables with respect to gauge transformations are:

\begin{eqnarray}
\{\underline{G}(\Lambda),U_\eta(\lambda,v)\}=0\nonumber\\
\{\underline{G}(\Lambda),U_\Theta(\delta,v)\}=-\frac{1}{2}ie\Lambda\delta\Theta(x_v)U_\Theta(\delta,v),\nonumber
\end{eqnarray}

which means that $U_\eta(\lambda,v)$ is gauge invariant and the transformation rule for $U_\Theta(\delta,v)$ is

\begin{eqnarray}
U_\Theta(\delta,v)\mapsto U_\Theta(\delta,v)U_\Theta(\frac{ae\Lambda\delta}{2},v)^{-1}.
\end{eqnarray}

The construction of the phase space is completely identical to the construction in case A, the only difference 
is that one has to replace the old variables with the new ones. To avoid confusion, 
$\mathcal{H}^{new}_{sc}=\mathcal{H}(U_\eta)\bigotimes\mathcal{H}(U_\Theta)$ will stand for the new phase space, 

\begin{eqnarray}
|\underline{\lambda}^\eta\rangle_\gamma=\prod_{k=1}^N U_\eta(\lambda_k^\eta,v_k)
\end{eqnarray}

will label an element of $\mathcal{H}(U_\eta)$ and 

\begin{eqnarray}
|\underline{\delta}^\Theta\rangle_\gamma=\prod_{k=1}^N U_\Theta(\delta_k^\Theta,v_k)
\end{eqnarray}

will be an element of $\mathcal{H}(U_\Theta)$. The action of these operators are completely the same as in case A:

\begin{eqnarray}
\hat{\Pi}_\eta(B)|\underline{\lambda}^\eta\rangle_\gamma = -\frac{\hbar}{2}\sum_{v_j\in B}\lambda_j^\eta|\underline{\lambda}^\eta\rangle_\gamma\\
\hat{U}_\eta(\lambda,v)|\underline{\lambda}^\eta\rangle_\gamma = |\underline{\lambda'}^\eta\rangle_\gamma\nonumber\\
\lambda^{,\eta}_i=\lambda^\eta_i+\delta_{v,v_i}\lambda
\end{eqnarray}

\subsection{Regularisation}

In order to quantize this system, one first has to rewrite the Hamiltonian in terms of the variables defined 
in the previous section - this is called the regularisation procedure. The key observation is (\cite{qsd5}) that if 
the gravitational field is dynamical, one can construct a well defined, diffeomorphism covariant Hamiltonian operator. 
In the article mentioned above the reader will find the detailed analysis of the gravitational, Yang-Mills, scalar 
and fermion fields. Since the method is quite lengthy, we are going to concentrate only on those terms that are different 
to the ones mentioned above. Specifically these are the terms that contain the scalar field. We will deal with the two 
kinds of description (the original case with $\Phi$ and $\Phi^*$ and the case with new variables $\eta$ and $\Theta$) 
separately.\\ \\
{\bf Case A:} 

Although later we will use the formulas involving $\eta$ and $\Theta$ we shall provide the regularisation of the original 
Hamiltonian, since it has some non trivial steps. The potential term is the simplest: since 
$\Phi\Phi^*=\Re(\Phi)^2+\Im(\Phi)^2$ and $\Re(\Phi)=\arccos\Big(\frac{U(\lambda,v)+U^{-1}(\lambda,v)}{2}\Big)$, we can write 
this (using the notations of \cite{qsd5}) in the following form:

\begin{eqnarray}
\hspace{-2cm}\hat{H}_{pot}=\frac{1}{4}\sum_v N(v)\mu\hat{V}\times\nonumber\\
\hspace{-2cm}\times\left[\frac{1}{\lambda^2}\arccos\left(\frac{U(\lambda,v)+U^{-1}(\lambda,v)}{2}\right)^2+\frac{1}{\delta^2}\arccos\left(\frac{\bar{U}(\delta,v)+\bar{U}^{-1}(\delta,v)}{2}\right)^2-a^2\right]^2\label{pot}
\end{eqnarray}

One may wonder why we used the arccos function instead of e.g. the logarithm. The main reason is that since 
spontaneous symmetry breaking requires the ground state of the potential, we are \underline{forced} to regularize 
the potential term to be self-adjoint. It is easy to see that the above operator is self-adjoint, but this would 
not be the case if we used the logarithm function. Of course there are still ambiguities in the regularisation, but 
this certainly narrows down the possibilities.\\
In a similar fashion, one replaces $\pi\pi^*=\Re(\pi)^2+\Im(\pi)^2$ in the kinetic term to obtain

\begin{eqnarray}
\hat{H}_P=\frac{1}{2}\sum_v N(v)\frac{X(v)^2+\bar{X}(v)^2}{E(v)^2}\hat{G}_1(v),
\end{eqnarray}

where $X(v)$ and $\bar{X}(v)$ are the invariant vector fields on U(1) and $\hat{G}_1(v)$ contains only gravitational 
variables and it is the same as in \cite{qsd5}:

\begin{eqnarray}
\hat{G}_1(v) &=& \frac{8}{81m^2\bar{h}^4\kappa^6}\sum_{v(\Delta)=v(\Delta`)=v}\epsilon^{IJK}\epsilon^{LMN}\epsilon_{ijk}\epsilon_{lmn}\times\nonumber\\
&\times& \hat{Q}^i_{s_I(\Delta)}(v,\frac{1}{2})\hat{Q}^j_{s_J(\Delta)}(v,\frac{1}{2})\hat{Q}^k_{s_K(\Delta)}(v,\frac{1}{2})\times\nonumber\\
&\times& \hat{Q}^l_{s_L(\Delta`)}(v,\frac{1}{2})\hat{Q}^m_{s_M(\Delta`)}(v,\frac{1}{2})\hat{Q}^n_{s_N(\Delta`)}(v,\frac{1}{2}),
\end{eqnarray}

where $\hat{Q}^k_{e}(v,r)=tr(\tau_k h_e[h^{-1}_e,\hat{V}(v)^r])$, $h_e$ being the holonomy of the Ashtekar connection along 
edge e, $\hat{V}$ is the volume operator and $\tau_k$ are the generators of SU(2).\\
The derivative term needs a more careful treatment. First we have to rewrite it in terms of $\Re(\Phi)$ and $\Im(\Phi)$

\begin{eqnarray}
D_a\Phi (D_b\Phi)^*=(\partial_a+ieA_a)(\Re(\Phi)+i\Im(\Phi))(\partial_b-ieA_b)(\Re(\Phi)-i\Im(\Phi))\nonumber
\end{eqnarray}

From this we can see that we need to regularize the expression $(\partial_a\pm ieA_a)\Re(\Phi)$. This is quite similar to 
the derivative term $\partial_a\Phi\pm A_a$ in \cite{Proca}, the only difference is that we have a $iA_a\Re(\Phi)$ term 
instead of $A_a$. Though this seems a minor change, it turns out that the regularized expression for this covariant 
derivative is more complicated, which is due to the fact that it contains the multiplication of the two fields. We can 
overcome this difficulty by doing the regularisation in a step-by-step way. First we note that for small 
$\Delta t$ $$h_s=1+ie\Delta t\dot{s}^a A_a+o(\Delta t^2)$$ for an edge s. This means that (v is the beginning of the edge s)
$$(h_s-1)\arccos\Big(\frac{U(\lambda,v)+U(\lambda,v)^{-1}}{2}\Big)=ie\lambda\Delta t\dot{s}^a A_a\Re(\Phi)+o(\Delta t^2),$$
so if we take into account that $$U(\lambda,s(\Delta t))=1+i\lambda(\Re(\Phi)+\Delta t\dot{s}^a\partial_a\Re(\Phi))+o(\Delta t^2),$$
we arrive to the following regularized expression:

\begin{eqnarray}
U(\lambda,s(\Delta t))[1+i(h_s-1)\arccos\Big(\frac{U(\lambda,v)+U(\lambda,v)^{-1}}{2}\Big)]U(\lambda,v)^{-1}=\nonumber\\
=[1+i\lambda(\Re(\Phi)+\Delta t\dot{s}^a\partial_a\Re(\Phi))](1-e\lambda\Delta t\dot{s}^a A_a\Re(\Phi))[1-i\lambda\Re(\Phi)]+o(\Delta t^2)=\nonumber\\
=1+i\lambda\Delta t\dot{s}^a(\partial_a\Re(\Phi))+ieA_a\Re(\Phi)))+o(\Delta t^2).
\end{eqnarray}

We obtain the same result for $(\partial_a+ieA_a)\Im(\Phi)$ if we replace $U$ with $\bar{U}$. Also the term 
$(\partial_a-ieA_a)\Re(\Phi)$ is obtained by replacing $h_s$ with $h^{-1}_s$. To simplify the result let us introduce 
a notation:

\begin{eqnarray}
W(v,s,\lambda)=\frac{1}{\lambda}U(\lambda,s(\Delta t))[1+i(h_s-1)\arccos\Big(\frac{U(\lambda,v)+U(\lambda,v)^{-1}}{2}\Big)]U(\lambda,v)^{-1}-\frac{1}{\lambda}\nonumber\\
\bar{W}(v,s,\delta)=\frac{1}{\delta}\bar{U}(\delta,s(\Delta t))[1+i(h_s-1)\arccos\Big(\frac{\bar{U}(\delta,v)+\bar{U}(\delta,v)^{-1}}{2}\Big)]\bar{U}(\delta,v)^{-1}-\frac{1}{\delta}.\nonumber
\end{eqnarray}

Using the fact that $h_s^{-1}=h_{s^{-1}},$ the regulated expressions are the following:

\begin{eqnarray}
W(v,s,\lambda) &=& i\Delta t\dot{s}^a(\partial_a\Re(\Phi))+ieA_a\Re(\Phi)))+o(\Delta t^2)\nonumber\\
W(v,s^{-1},\lambda) &=& i\Delta t\dot{s}^a(\partial_a\Re(\Phi))-ieA_a\Re(\Phi)))+o(\Delta t^2)\nonumber\\
\bar{W}(v,s,\delta) &=& i\Delta t\dot{s}^a(\partial_a\Im(\Phi))+ieA_a\Im(\Phi)))+o(\Delta t^2)\nonumber\\
\bar{W}(v,s^{-1},\delta) &=& i\Delta t\dot{s}^a(\partial_a\Im(\Phi))-ieA_a\Im(\Phi)))+o(\Delta t^2)\nonumber
\end{eqnarray}

This way the derivative term will be the limit of

\begin{eqnarray}
\hspace{-2cm}\hat{H}_{der}=\frac{1}{2}\sum_v\frac{N(v)}{E(v)^2}\times\nonumber\\
\hspace{-2cm}\times\sum_{v(\Delta)=v(\Delta`)=v}[W_\Delta(v,s_n,\lambda)+i\bar{W}_\Delta(v,s_n,\delta)][W_{\Delta'}(v,s^{-1}_r,\lambda)-i\bar{W}_{\Delta'}(v,s^{-1}_r,\delta)]\hat{G}_2^{nr}(v),\nonumber
\end{eqnarray}

where 

\begin{eqnarray}
\hspace{-1cm}\hat{G}_2^{nr}(v)=\frac{1}{2\bar{h}^4\kappa^4}\Big(\frac{4}{3}\Big)^6\epsilon^{ijk}\epsilon^{ilm}\epsilon_{npq}\epsilon_{rst}\hat{Q}^j_{s_p(\Delta)}(v,\frac{3}{4})\hat{Q}^k_{s_q(\Delta)}(v,\frac{3}{4})\hat{Q}^l_{s_s(\Delta`)}(v,\frac{3}{4})\hat{Q}^m_{s_t(\Delta`)}(v,\frac{3}{4})\nonumber
\end{eqnarray}

and the $\Delta$ and $\Delta'$ subscripts represent the tetrahedra where the holonomies and pointholonomies should be 
calculated.\\ \\
{\bf Case B:} 

In this case one should be careful since the two scalar fields do not appear in a symmetric way. For instance, 
the potential term contains only $\eta$, so one simply replaces 
$\eta=\frac{1}{\lambda}\arccos\Big(\frac{U_\eta(\lambda,v)+U^{-1}_\eta(\lambda,v)}{2}\Big)$ to obtain

\begin{eqnarray}
\hat{H}_{pot}=\frac{1}{4}\sum_v N(v)\mu\hat{V}\frac{1}{\lambda^2}\arccos\Big(\frac{U_\eta(\lambda,v)+U^{-1}_\eta(\lambda,v)}{2}\Big)^2\times\nonumber\\
\times\Big[\frac{1}{\lambda}\arccos\Big(\frac{U_\eta(\lambda,v)+U^{-1}_\eta(\lambda,v)}{2}\Big)+2a\Big]^2
\end{eqnarray}

The terms containing $\pi_\eta$ and $\pi_\Theta$ can be treated in the same way as in the previous case, one 
just has to be careful since the later contains the expression $\frac{1}{(a+\eta)^2}$. But it is easy to see 
that if one carries out the regularisation procedure as in \cite{qsd5} the only difference will be a term which 
is the above fraction expressed with the variables $U_\eta$. Thus the result for the two kinetic terms will be 

\begin{eqnarray}
\hat{H}_P=\frac{1}{2}\sum_v N(v)\frac{X_\eta(v)^2+\Big[\frac{1}{\lambda}\arccos\Big(\frac{U_\eta (\lambda,v)+U^{-1}_\eta (\lambda,v)}{2}\Big)+a\Big]^{-2} X_\Theta (v)^2}{E(v)^2}\hat{G}_1(v).
\end{eqnarray}

The derivative terms for the two scalar fields are also different. In the case of $\eta$, one only needs the 
expression (for small $\Delta t$):

\begin{eqnarray}
U_\eta(\lambda,s(\Delta t))U^{-1}_\eta(\lambda,v)-1=i\lambda\Delta t\dot{s}^a\partial_a\eta+o(\Delta t^2),
\end{eqnarray}

thus it has contribution to the Hamiltonian
\begin{eqnarray}
\hat{H}_{der}(\eta)=\frac{1}{2}\sum_v\frac{N(v)}{E(v)^2}\sum_{v(\Delta)=v(\Delta`)=v}\frac{1}{\lambda^2}[U_\eta(\lambda,s_n(\Delta))U^{-1}_\eta(\lambda,v)-1]\times\nonumber\\
\times[U_\eta(\lambda,s_r(\Delta'))U^{-1}_\eta(\lambda,v)-1]\hat{G}_2^{nr}(v)
\end{eqnarray}

For the field $\Theta$ we remind the reader that in \cite{Proca} the same kind of coupling appeared between 
the scalar field and the Maxwell field. Thus we may use the approximation mentioned there:

\begin{eqnarray}
U_\Theta(\delta,s(\Delta t))\underline{h}_s U_\Theta(\delta,v)^{-1}-1=i\delta\Delta t\dot{s}^b(\frac{\partial_b\Theta}{a}+e\underline{A}_b)
\end{eqnarray}

Treating the term $(a+\eta)^2$ as before, the regulated expression of this term will be

\begin{eqnarray}
\hspace{-2cm}\hat{H}_{der}(\Theta)=\frac{1}{2}\sum_v\frac{N(v)}{E(v)^2}\Big[\frac{1}{\lambda}\arccos\Big(\frac{U_\eta (\lambda,v)+U^{-1}_\eta (\lambda,v)}{2}\Big)+a\Big]^{2}\sum_{v(\Delta)=v(\Delta`)=v}\nonumber\\
\hspace{-2cm}\frac{1}{\delta^2}[U_\Theta(\delta,s_n(\Delta))\underline{h}_{s_n}U_\Theta(\delta,v)^{-1}-1][U_\Theta(\delta,s_r(\Delta'))\underline{h}_{s_r}U_\Theta(\delta,v)^{-1}-1]\hat{G}_2^{nr}(v)
\end{eqnarray}

\section{New basis}

In contrast to the Proca field, the mass here is represented by an operator, namely

\begin{eqnarray}
\hat{m}(v)=e\Big[\frac{1}{\lambda}\arccos\Big(\frac{U_\eta(\lambda,v)+U^{-1}_\eta(\lambda,v)}{2}\Big)+a\Big].
\end{eqnarray}

Since we want to compare the two theories, it would be useful to work in a basis where $U_\eta(\lambda,v)$ is 
diagonal and this is what we are going to do in this section. 

\subsection{The spectrum of $U_\eta(\lambda,v)$}

Let $|\phi\rangle:=|\lambda^\eta_{v_1},\dots,\lambda^\eta_{v_N}\rangle$ be a base element for a $\gamma$ graph 
which has N vertices The action of $U_\eta(\lambda,v)$ on this state is

\begin{eqnarray}
U_\eta(\lambda,v)|\phi\rangle=|\lambda^\eta_{v_1},\dots,\lambda^\eta_{v_k}+\lambda,\dots,\lambda^\eta_{v_N}\rangle\delta(v,v_k)
\end{eqnarray}

This action suggests that we should look for eigenstates in the form

\begin{eqnarray}
|\Lambda^\eta(\lambda,v),\underline{\lambda}^\eta\rangle:=\sum_{i=-\infty}^\infty\Big(\frac{U_\eta(\lambda,v)}{\Lambda^\eta(\lambda,v)}\Big)^i|\underline{\lambda}^\eta\rangle,
\end{eqnarray}

where $|\underline{\lambda}^\eta\rangle$ is an arbitrary state and $\Lambda^\eta(\lambda,v)$ is a (yet) arbitrary 
number (this will be the eigenvalue for a given $\lambda$ at a vertex v). It is easy to verify that 

\begin{eqnarray}
U_\eta(\lambda,v)|\Lambda^\eta(\lambda,v),\underline{\lambda}^\eta\rangle=\Lambda(\lambda,v)|\Lambda^\eta(\lambda,v),\underline{\lambda}^\eta\rangle.
\end{eqnarray}

We shall call these one vertex eigenstates because $|\Lambda^\eta(\lambda,v),\underline{\lambda}^\eta\rangle$ is the 
eigenstate of only those $U_\eta(\lambda,v')$ where v'=v.Since $U_\eta(\lambda,v)$ is unitary, we can write 
$\Lambda^\eta(\lambda,v)$ in the following form: $\Lambda^\eta(\lambda,v)=\exp(i\Delta^\eta(\lambda,v))$, 
where $\Delta^\eta$ is real.In fact, since $U_\eta(0,v)=\hat{1}$ and 
$U_\eta(\lambda_1,v)U_\eta(\lambda_2,v)=U_\eta(\lambda_1+\lambda_2,v)$, we obtain that $\Delta^\eta(\lambda,v)$ is of 
the form $\Delta^\eta(\lambda,v)=\Gamma^\eta(v)\lambda$. In summary, the spectrum of $U_\eta(\lambda,v)$ is of the 
form $\exp(i\lambda\Gamma^\eta(v))$, so instead of $\Lambda^\eta(\lambda,v)$ we shall use $\Gamma^\eta(v)$. We can select 
an orthonormal basis from these eigenstates in the following way. Note that if there exists an integer n such that 
$|\underline{\lambda}^\eta_1\rangle=U(\lambda,v)^n|\underline{\lambda}^\eta_2\rangle$ then 
$|\Gamma^\eta(v),\underline{\lambda}^\eta_1\rangle=e^{in\lambda\Gamma^\eta(v)}|\Gamma^\eta(v),\underline{\lambda}^\eta_2\rangle$. 
Because of this let us restrict ourselves to those $|\underline{\lambda}^\eta\rangle$ that satisfy the condition 
$0\leq\lambda^\eta_v\langle\lambda$. Further more if we restrict the values of $\Gamma^\eta(v)$ so that 
$0\leq\Gamma^\eta(v)\langle\frac{2\pi}{\lambda}$, these states will form a complete orthonormal basis in the sense

\begin{eqnarray}
\langle\Gamma^\eta_1(v),\underline{\lambda}_1|\Gamma^\eta_2(v),\underline{\lambda}_2\rangle=\nonumber\\
=\sum_{k=-\infty}^\infty\sum_{j=-\infty}^\infty\langle\lambda^{\eta,1}_{v_1},\dots,\lambda^{\eta,1}_v+k\lambda,\dots,\lambda^{\eta,1}_{v_N}|\lambda^{\eta,2}_{v_1},\dots,\lambda^{\eta,2}_v+j\lambda,\dots,\lambda^{\eta,2}_{v_M}\rangle\times\nonumber\\
\times\exp(i\lambda(j\Gamma^\eta_2(v)-k\Gamma^\eta_1(v)))=\delta(\underline{\lambda}^\eta_1,\underline{\lambda}^\eta_2)\sum_{k=-\infty}^\infty \exp(i\lambda k(\Gamma^\eta_2(v)-\Gamma^\eta_1(v)))=\nonumber\\
=\delta(\underline{\lambda}^\eta_1,\underline{\lambda}^\eta_2)\delta(\Gamma^\eta_2(v)-\Gamma^\eta_1(v))),
\end{eqnarray}

where $\delta(\underline{\lambda}^\eta_1,\underline{\lambda}^\eta_2)=\delta(\lambda^{(1)\eta}_1-\lambda^{(1)\eta}_2)\dots\delta(\lambda^{(N)\eta}_1-\lambda^{(N)\eta}_2)$. 
To see that this is a {\it complete} orthonormal basis, one only has to check whether each original basis element 
can be expressed as the linear combination of the eigenstates. Let us suppose then that there exist complex numbers 
$C_{\underline{\lambda}}(\Gamma^\eta(v))$ such that

\begin{eqnarray}
\sum_{\underline{\lambda}^\eta}\int d\Gamma^\eta(v)C_{\underline{\lambda}^\eta}(\Gamma^\eta(v))|\Gamma^\eta(v),\underline{\lambda}^\eta\rangle=|\underline{\lambda}^{\eta,}\rangle
\end{eqnarray}

for each $|\underline{\lambda}^{\eta,}\rangle$. Because of orthogonality we obtain for the coefficients the following:

\begin{eqnarray}
\hspace{-2cm}C_{\underline{\lambda}^\eta}(\Gamma^\eta(v))=\langle\Gamma^\eta(v),\underline{\lambda}^\eta|\underline{\lambda}^{\eta,}\rangle=\langle\underline{\lambda}^\eta|\sum_{k=-\infty}^{\infty}\exp(-ik\lambda\Gamma^\eta(v))U_\eta(\lambda,v)^k|\underline{\lambda}^{\eta,}\rangle
\end{eqnarray}

Now if for a $|\underline{\lambda}^\eta\rangle$ there exists an integer n such that 
$|\underline{\lambda}^\eta\rangle=U(\lambda,v)^n|\underline{\lambda}^{\eta,}\rangle$ then the corresponding coefficient will be 
$$C_{\underline{\lambda}^\eta}(\Gamma^\eta(v))=\exp(-in\lambda\Gamma^\eta(v)),$$ otherwise it is zero. It is easy to see that 
this correspondence is unique and since the original basis is complete, we verified our statement.\\ \\
We define the graph eigenstate in a similar fashion. Let $\gamma$ be a graph and $|\underline{\lambda}^\eta\rangle$ be an 
arbitrary state on that graph. For each vertex let $\Gamma^\eta(v_i)\ i=1\dots N$ be a real number satisfying 
$0\leq\Gamma^\eta(v_i)\langle\frac{2\pi}{\lambda}$. Then the graph eigenstate will be the following:

\begin{eqnarray}
\hspace{-2cm}|\underline{\Gamma}^\eta,\underline{\lambda}^\eta\rangle:=\sum_{k_1=-\infty}^\infty\dots\sum_{k_N=-\infty}^\infty (e^{ik_1\lambda\Gamma^\eta(v_1)}U_\eta(\lambda,v_1)^{k_1})\dots (e^{ik_N\lambda\Gamma^\eta(v_N)}U_\eta(\lambda,v_N)^{k_N})|\underline{\lambda}^\eta\rangle
\end{eqnarray}

Using the results obtained for the one vertex eigenstates we can find an orthonormal basis in the case of the 
graph eigenstates: if $0\leq\lambda^\eta_v\langle\lambda$ for all v we get

\begin{eqnarray}
\hspace{-2cm}\langle\underline{\Gamma}_1^\eta,\underline{\lambda}_1^\eta|\underline{\Gamma}_2^\eta,\underline{\lambda}_2^\eta\rangle=\delta(\underline{\lambda}_1^\eta,\underline{\lambda}_2^\eta)\prod_{k=1}^N\delta(\Gamma^\eta_1(v_1)-\Gamma^\eta_2(v_1))\dots\delta(\Gamma^\eta_1(v_N)-\Gamma^\eta_2(v_N))
\end{eqnarray}

Also we can express any state in terms of graph eigenstates with the help of the following expression:

\begin{eqnarray}
\sum_{\underline{\lambda}^\eta}\int d\underline{\Gamma}^\eta C_{\underline{\lambda}^\eta}(\underline{\Gamma}^\eta)|\underline{\Gamma}^\eta,\underline{\lambda}^\eta\rangle=|\underline{\lambda}^{\eta,}\rangle
\end{eqnarray}

where $$\int d\underline{\Gamma}^\eta=\int d\Gamma^\eta(v_1)\dots\int d\Gamma^\eta(v_N).$$ What remains is the 
action of the momentum operators on an eigenstate. This is easy because of the following:

\begin{eqnarray}
X(v)|\Gamma^\eta(v'),\underline{\lambda}\rangle=X(v)\sum_{k=-\infty}^\infty\Big(\frac{U_\eta(\lambda,v')}{\exp(i\lambda\Gamma^\eta(v'))}\Big)^k|\underline{\lambda}^\eta\rangle=\nonumber\\
=\delta_{v,v_k}\lambda_k|\Gamma^\eta(v'),\underline{\lambda}\rangle-i\delta_{v,v'}\sum_{k=-\infty}^\infty ik\lambda\Big(\frac{U_\eta(\lambda,v')}{\exp(i\lambda\Gamma^\eta(v'))}\Big)^k|\underline{\lambda}^\eta\rangle=\nonumber\\
=(\delta_{v,v_k}\lambda_k+i\delta_{v,v'}\frac{\delta}{\delta\Gamma(v')})|\Gamma^\eta(v'),\underline{\lambda}\rangle\label{pnewbasis}
\end{eqnarray}

With a completely similar analysis one can show that

\begin{eqnarray}X(v)
|\underline{\Gamma}^\eta,\underline{\lambda}\rangle=(\delta_{v,v_k}\lambda_k+i\delta_{v,v'}\frac{\delta}{\delta\Gamma(v')})|\underline{\Gamma}^\eta,\underline{\lambda}\rangle
\end{eqnarray}

\section{Solution to the constraints}

In \cite{Proca} we sketched how one could solve the constraints of the theory. In this section we will follow the 
same procedures mentioned there - especially in the case of the scalar constraint.  This method can also be used 
in this  case, with one difference, namely that for the fields $\eta$ and $\Theta$ we do not work in the usual basis, 
rather in the Fock-space. Since for the other fields the algorithm remains the same, we will concentrate only on 
the scalar fields. Solving the diffeomorphism- and gauge constraints will be rather simple, so we start with them. 
Then - in order to simplify things - we will introduce a compact notation where we separate the scalar fields from 
the others, which is described in Appendix A. This is motivated by the fact that the scalar constraint is quite 
complicated, but with the new notation the structure of the equation will be much easier to examine.\\
Let us start with the diffeomorphism constraint. As it was pointed out in \cite{revt} the infinitesimal generator of 
the diffeomorphism constraint cannot be implemented in the quantum theory, thus the techniques used to solve the 
Gauss- or scalar constraint cannot be applied here. The strategy is to use group averaging to solve the constraint, 
which can be generalized to the case where matter fields also appear (see \cite{qsd5} for details). Since these are 
applied only to graphs not to the labels means that it is independent whether we use the Fock-space or the dust network space.\\
The gravitational Gauss-constraint is the same as in \cite{revt}, so we can solve it by restricting ourselves to 
gauge invariant spin network states.\\
The U(1) Gauss-constraint contains variables of the electromagnetic field and the scalar field $\Theta$ so we 
analyze it in detail. The (smeared) integrated constraint

\begin{eqnarray}
\int_\sigma\underline{G}\Lambda=\int_\sigma\Lambda(D_a \underline{E}_a-ae\pi_\Theta)
\end{eqnarray}

can be regulated in the following way: Let us look for solutions in the form 

\begin{eqnarray}
\Psi=\sum_{s,f,\underline{\lambda},\underline{\bar{\delta}}}\int d\underline{\Gamma}\int d\underline{\bar{\Gamma}} C_{s,f,\underline{\lambda},\underline{\bar{\delta}}}(\underline{\Gamma},\underline{\bar{\Gamma}})\langle\underline{s}|\langle\underline{f}|\langle\underline{\Gamma},\underline{\lambda}|\langle\underline{\bar{\Gamma}},\underline{\bar{\delta}}|.\label{solution}
\end{eqnarray}

It can be verified that the quantum version of the above constraint is the following:

\begin{eqnarray}
\langle\Psi|\sum_v\Lambda_v[\sum_{e\cap v=v}l_e-(\delta_v+i\frac{\delta}{\delta\bar{\Gamma}(v)})]|\Phi\rangle=0
\end{eqnarray}

for all spin color network state $|\Phi\rangle$. Here $l_e$ is the integer on the edge e (this comes from the flux network). 
Since $\Lambda_v$ is arbitrary the above equation is equivalent to

\begin{eqnarray}
\int d\bar{\Gamma}(v)' C_{s,f,\underline{\lambda},\underline{\bar{\delta}}'}(\underline{\Gamma},\underline{\bar{\Gamma}}')\langle\underline{\bar{\Gamma}}',\underline{\bar{\delta}}'|\sum_{e\cap v=v}l_e-(\delta_v+i\frac{\delta}{\delta\bar{\Gamma}(v)})|\underline{\bar{\Gamma}},\underline{\bar{\delta}}\rangle=0,
\end{eqnarray}

where we inserted \eref{solution} to the constraint equation and used orthogonality of spin color network states. 
Now after partial integration we obtain a (functional) differential equation on the coefficients 
$C_{s,f,\underline{\lambda},\underline{\bar{\delta}}}(\underline{\Gamma},\underline{\bar{\Gamma}})$:

\begin{eqnarray}
[\sum_{e\cap v=v}l_e-(\delta_v-i\frac{\delta}{\delta\bar{\Gamma}(v)})]C_{s,f,\underline{\lambda},\underline{\bar{\delta}}}(\underline{\Gamma},\underline{\bar{\Gamma}})=0.\label{eqgauge}
\end{eqnarray}

Since we have a similar equation for all v, the solution to this constraint is:

\begin{eqnarray}
C_{s,f,\underline{\lambda},\underline{\bar{\delta}}}(\underline{\Gamma},\underline{\bar{\Gamma}})=C_{s,f,\underline{\lambda},\underline{\bar{\delta}}}(\underline{\Gamma})\prod_v\exp[-i(\sum_{e\cap v=v}l_e-\delta_v)\bar{\Gamma}(v)],\label{resgauge}
\end{eqnarray}

where the coefficients $C_{s,f,\underline{\lambda},\underline{\bar{\delta}}}(\underline{\Gamma})$ are arbitrary.\\
What remains is the scalar constraint. If we look at the Hamiltonian, it is clear that the constraint equation 
will be a differential equation with respect to the variable $\Gamma$. First we write down this equation. The condition 
we have to solve is

\begin{eqnarray}
\langle\Psi|\hat{H}|\phi\rangle=0
\end{eqnarray}

for arbitrary $|\phi\rangle$. Again we can say that the support of N is at only one vertex v. Substituting 
\eref{solution} into the above equation we obtain

\begin{eqnarray}
\sum_{s',f',\underline{\lambda}',\underline{\bar{\delta}}'}\int d\Gamma(v)'\int d\bar{\Gamma(v)}' C_{s',f',\underline{\lambda}',\underline{\bar{\delta}}'}(\underline{\Gamma}',\underline{\bar{\Gamma}}')\times\nonumber\\
\times \langle s'|\langle f'|\langle\underline{\Gamma}',\underline{\lambda}'|\langle\underline{\bar{\Gamma}}',\underline{\bar{\delta}}'|\hat{H}|\underline{\bar{\Gamma}},\underline{\bar{\delta}}\rangle|\underline{\Gamma},\underline{\lambda}\rangle|f\rangle|s\rangle=0\nonumber
\end{eqnarray}

In \cite{Proca} we have shown a method (generalizing the results of \cite{qsd}) which simplified the above 
equation by turning it into a finite number of equations. The main idea is that we take a basis element 
$|s\rangle|f\rangle$ (discarding the scalar field for the moment) and we create a set $S^{(1)}$ containing 
basis elements appearing in $\hat{H}|s\rangle|f\rangle$. We continue this procedure and construct $S^{(n)}$ recursively 
from $S^{(n-1)}$. After this we search for solutions of the form
$$\langle\Psi|=\sum_i\sum_{\langle f|\langle s|\in S^{(i)}}C^i_{sf}\langle s|\langle f|,$$ where s is a basis element. 
One can show that if we substitute this into the constraint equation we arrive to a finite number of conditions 
(details can be found in the mentioned articles). Now we apply these results to the gravitational and 
electromagnetic fields and we arrive to a finite system of linear differential equations. Since we concentrate 
only on the scalar field $|\Gamma,\lambda\rangle$ at the moment and the Hamiltonian contains several terms, we 
shall calculate each term separately and introduce a compact notation. This notation is introduced in appendix A 
where the reader will also find the terms of the scalar constraint. The conclusion is that the scalar constraint 
is actually a system of linear differential equations of second order:

\begin{eqnarray}
\sum_{I'}H^P_{I'I}(\lambda_v-i\frac{\delta}{\delta\Gamma(v)})\tilde{C}_{I'}(\underline{\Gamma})=-\sum_{I'}H_{I'I}(\underline{\Gamma})\tilde{C}_{I'}(\underline{\Gamma})
\end{eqnarray}

where

\begin{eqnarray}
H_{I'I}(\underline{\Gamma})=\nonumber\\
=\frac{L(v)^2 H^P_{I'I}(v)}{(\Gamma(v)+a)^2}+H^{G+YM}_{I'I}(v)+H^{der1}_{I'I}(\Gamma(v)+a)^2+H^{pot}_{I'I}(v)\Gamma(v)^2(\Gamma(v)+2a)^2+\nonumber\\
+H^{A}_{I'I}(v)\exp(-2i\lambda\Gamma(v))-H^{B}_{I'I}(v)\exp(-i\lambda\Gamma(v))+H^{C}_{I'I}(v)\label{hamdiff}
\end{eqnarray}

To simplify this term we look for solutions of the form 
$$C_I(\underline{\Gamma}):=\tilde{C}_I(\underline{\Gamma})\exp(-i\prod_v\lambda_v\Gamma(v)),$$ since 
$$(\lambda_v-i\frac{\delta}{\delta\Gamma(v)})C_I(\underline{\Gamma})=-i\Big(\frac{\delta}{\delta\Gamma(v)}\tilde{C}_I(\underline{\Gamma})\Big)\exp(-i\prod_v\lambda_v\Gamma(v)).$$ 
The other terms will also contain a factor $\exp(-i\prod_v\lambda_v\Gamma(v))$ so this drops out of the 
differential equation, leaving us with the following formula for $\tilde{C}_{I'}(\underline{\Gamma}):$

\begin{eqnarray}
\sum_{I'}H^P_{I'I}\frac{\delta^2}{\delta\Gamma(v)^2}\tilde{C}_{I'}(\underline{\Gamma})=\sum_{I'}H_{I'I}(\underline{\Gamma})\tilde{C}_{I'}(\underline{\Gamma})
\end{eqnarray}

\subsection{Solving the scalar constraint}

This system of linear differential equations can be solved using the method we shown in Appendix B if the 
matrix $H^P_{I'I}$ is invertible. If it is not invertible then let us diagonalize the left hand side, i.e. 
find a unitary ${\bf U}$ such that ${\bf U}{\bf H}^P {\bf U}^{-1}=diag(k_1,\dots,k_N)$ where $k_1,\dots,k_N$ are 
the eigenvalues of ${\bf H}^P$. Let us order the eigenvalues in a way that $k_1,\dots,k_M$ ($M\langle N$) be all 
the zero eigenvalues. This means that the first M equation in this case is not a differential equation but only 
an algebraic equation. Since in this case the left hand side is zero, the right hand side is zero if and only 
$\sum_{I'=1}^M H^{,Q}_{I'I}(v)\tilde{C}_{I'}=0$ etc. for all matrices appearing in 
$H_{I'I}(\underline{\Gamma})$ ($H^{,Q}_{I'I}(v)=({\bf U}{\bf H}^Q(v) {\bf U}^{-1})_{I'I}$ etc.), which means that 
after solving the algebraic equations we again arrive to a system of linear differential equations but with an 
invertible matrix on the left hand side. So from now on we consider $H^P_{I'I}$ to be invertible.\\
To have a correct solution we must specify the initial condition on $\tilde{C}_{I'}(\underline{\Gamma})$ and 
$\frac{\partial}{\partial\Gamma(v)}\tilde{C}_{I'}(\underline{\Gamma})$. The fact that $\Gamma(v)=-a$ can be 
interpreted as the disappearance of the field $\eta$ implies that

\begin{eqnarray}
[\sum_{I'}H_{I'I}(\underline{\Gamma})\tilde{C}_{I'}(\underline{\Gamma})]_{|\Gamma(v)=-a}=0\label{init1}
\end{eqnarray}

be the first condition. With the same reasoning the second condition is that the momentum of the field should 
disappear. In this case (since all $\lambda_v$ are zero) we arrive to the condition

\begin{eqnarray}
[\frac{\partial}{\partial\Gamma(v)}\tilde{C}_{I'}(\underline{\Gamma})]_{|\Gamma(v)=-a}=0\label{init2}
\end{eqnarray}

Since $H_{I'I}$ has a complicated structure, the differential equation cannot be solved explicitly. 
However we can solve it in some special case.\\
First let us consider the case when $\Gamma(v)\approx -a$. In this case the system of differential 
equations takes the form

\begin{eqnarray}
\sum_{I'}H^P_{I'I}\frac{\delta^2}{\delta\Gamma(v)^2}\tilde{C}_{I'}(\underline{\Gamma})=\sum_{I'}\frac{L(v)^2 H^P_{I'I}(v)}{(\Gamma(v)+a)^2}\tilde{C}_{I'}(\underline{\Gamma})+\sum_{I'}H_{I'I}^{G+YM}\tilde{C}_{I'}(-a)
\end{eqnarray}

Now if we multiply both sides with $({\bf H}^P)^{-1}$ and define $b_I=(({\bf H}^P)^{-1}{\bf H}^{G+YM}\vec{C}(-a))_I$ we get

\begin{eqnarray}
\frac{\delta^2}{\delta\Gamma(v)^2}\tilde{C}_I(\underline{\Gamma})-\frac{L(v)^2}{(\Gamma(v)+a)^2}\tilde{C}_I(\underline{\Gamma})=b_I
\end{eqnarray}

The general solution of this differential equation is the following:

\begin{eqnarray}
\tilde{C}_I(\underline{\Gamma})=\frac{(\Gamma(v)+a)^2}{2-L(v)^2}b_I+C^1_I(\Gamma(v)+a)^{n_1}+C^2_I(\Gamma(v)+a)^{n_2},
\end{eqnarray}

where $$n^1_2=\frac{1\pm\sqrt{1+4L(v)^2}}{2}$$ and $C^1_I,C^2_I$ are arbitrary constants. From $L(v)^2\geq 0$ follows 
that $n_1\geq 1$ and $n_2\leq 0$, which means that if $L(v)\neq 0$ then $\tilde{C}_I(\underline{\Gamma})$ is singular 
in $\Gamma(v)=-a$.\\
Now let us consider the initial conditions. If $L(v)=0$ then 

\begin{eqnarray}
\tilde{C}_I(\underline{\Gamma})=\frac{(\Gamma(v)+a)^2}{2}b_I+C^1_I(\Gamma(v)+a)+C^2_I.
\end{eqnarray}

Substituting into \eref{init1} and \eref{init2} implies that $C^1_I=0$ and $b_I=0$. Further more from the 
definition of $b_I$ comes that $b_I=(({\bf H}^P)^{-1}{\bf H}^{G+YM}\vec{C}^2)_I$, so the solution is:

\begin{eqnarray}
\tilde{C}_I(\underline{\Gamma})=C^2_I,
\end{eqnarray}

where $C^2_I$ must satisfy the condition

\begin{eqnarray}
H^{G+YM}_{I'I}C^2_I=0.
\end{eqnarray}

This is not a surprising result since if $L(v)=0$ then substituting this into the constraints we obtain a 
theory completely equivalent to the electromagnetic field coupled to gravity. If we rewrite the scalar constraint 
of this theory in terms of the notation used in appendix A, we obtain the above condition.\\
What happens if $L(v)\neq 0$. In this case $C^2_I=0$ so that the solution does not become singular at $\Gamma=-a$. 
Substituting into \eref{init2} will yield the identity 0=0, so we must check \eref{init1}. For $L(v)=\pm 1$ this 
will be singular so in this case $C^1_I=0$ and only the first term survives, but it will be zero too. 
Thus in this case the solution near $\Gamma=-a$ is zero in first order. For $|L(v)|\rangle1$ the condition 
\eref{init1} is also an identity. But in this case $\tilde{C}_I(-a)=0$, so $b_I=0$, thus

\begin{eqnarray}
\tilde{C}_I(\underline{\Gamma})=C^1_I(\Gamma(v)+a)^{n_1}.
\end{eqnarray}

This solution tends rapidly to zero as $\Gamma\to -a$ (especially if $L(v)$ is large), so as we reach this limit, 
the amplitude of the solution coming from the $L(v)=0$ case will become significantly larger. In fact the larger 
$L(v)$ is, the amplitude becomes much smaller in this region. So we can say that if $\Gamma(v)+a\approx 0$ 
(which - as we will see later - can be interpreted as the mass is about zero) states which for which $L(v)=0$ have 
the highest probability while the larger $|L(v)|$, the smaller this probability will get.\\
These results show that in contrast to the Proca field, this theory provides us with the different amplitudes for 
different masses. However because the two theories are - in some aspect - very similar, it would be desirable to provide 
the solutions of this theory which can be identified as the solutions to the Proca field. The basic idea is very 
simple: we compare the two Hamiltonians. If we look at the matrix \eref{hamdiff} in our differential equation, in 
the case $\Gamma=0$ it will be the same as the Hamiltonian of the Proca field. So one just needs to imply the 
conditions

\begin{eqnarray}
\frac{\delta^2}{\delta\Gamma(v)^2}\tilde{C}_{I'}(\underline{\Gamma})=0\label{proccond1}\\
(\sum_{I'}H_{I'I}(\underline{\Gamma})\tilde{C}_{I'}(\underline{\Gamma}))_{|\Gamma=0}=0\label{proccond2}
\end{eqnarray}

The problem is that in this theory this will provide a distributional solution in the following sense. In the case of 
the Proca field the mass is fixed, which means that we are interested in solutions where $\Gamma$ is constant. 
But now we have a differential equation so $\Gamma$ is continuous. The way out of this is we say that in the 
interval $(-\epsilon,\epsilon)$ we solve \eref{proccond1}, and outside this interval $\tilde{C}_{I'}(\underline{\Gamma})$ 
is zero. The required solution will the limit $\epsilon\to 0$. The reason for this strange behavior is that the 
equation we gained looks not like the Proca, but the linear combination of all the Procas. 

\section{Mass}

In quantum field theory the mass is the coefficient of the term in the Hamiltonian which is quadratic in the boson 
field. However in this case we shall define the mass as an operator corresponding to the classical expression $\eta+a$. 
The reasons for us to do so are the following: First - as was shown at the end of section 2.2 - the expression $\eta+a$ 
corresponds exactly to the mass parameter of the Proca field (The term $(\eta+a)^2$ not only appears in front of the 
quadratic term of the bosonic field but also appears in the denometer of the kinetic term of the other scalar field). 
The second reason is that in this case we can simplify our analysis regarding the scalar-bozon interaction. This new 
interpretation - as we will see - gives a better understanding of the mass generation in the Hamiltonian framework. 
Note also that the substitution $\eta=0$ gives back the ``original'' mass.\\
Let $|\Psi\rangle:=\sum_{\underline{\lambda}^\eta}\int d\underline{\Gamma}^\eta C_{\underline{\lambda}^\eta}(\underline{\Gamma}^\eta)|\underline{\Gamma}^\eta,\underline{\lambda}^\eta\rangle$ be a solution of the constraints. Then we can define the 
``mass operator'' as

\begin{eqnarray}
\hspace{-2cm}\hat{m}|\Psi\rangle=\frac{1}{\lambda}\arccos\Big(\frac{U_\eta(\lambda,v)+U^{-1}_\eta(\lambda,v)}{2}\Big)|\Psi\rangle=\nonumber\\
=\sum_{\underline{\lambda}^\eta}\int d\underline{\Gamma}^\eta C_{\underline{\lambda}^\eta}(\underline{\Gamma}^\eta)\Gamma^\eta(v)|\underline{\Gamma}^\eta,\underline{\lambda}^\eta\rangle.
\end{eqnarray}

It is clear from the definition that this operator is self adjoint, thus it has real eigenvalues. Further more its 
spectrum is continuous. Its expectation value is 

\begin{eqnarray}
\hspace{-2cm}m(\Psi,v)=\sum_{\underline{\lambda}_1^\eta,\underline{\lambda}_2^\eta}\int d\underline{\Gamma}^\eta_1 d\underline{\Gamma}^\eta_2 C^*_{\underline{\lambda}^\eta}(\underline{\Gamma}^\eta_1)C_{\underline{\lambda}^\eta}(\underline{\Gamma}^\eta_2)e(\Gamma_2^\eta(v)+a)\langle\underline{\Gamma}^\eta_1,\underline{\lambda}_1^\eta|\underline{\Gamma}^\eta_2,\underline{\lambda}_2^\eta\rangle=\nonumber\\
=\sum_{\underline{\lambda}^\eta}\int d\underline{\Gamma}^\eta|C_{\underline{\lambda}^\eta}(\underline{\Gamma}^\eta)|^2 e(\Gamma^\eta(v)+a),
\end{eqnarray}

thus for a graph $\gamma$ we may define the mass as

\begin{eqnarray}
m(\Psi,\gamma)=\sum_v m(\Psi,v)=\sum_{\underline{\lambda}^\eta}\int d\underline{\Gamma}^\eta|C_{\underline{\lambda}^\eta}(\underline{\Gamma}^\eta)|^2\sum_v e^2(\Gamma^\eta(v)+a).
\end{eqnarray}

This means that $\Gamma(v)+a$ can be interpreted as ``mass in a vertex''. If we look at a state where all $\Gamma$ are zero - 
the ``vacuum'' (note that it is NOT the usual vacuum since we are not in the Fock space representation) - we obtain states 
with mass $ea$. But one may ask whether this is an observable or not. If one checks the commutator of the constraints and 
$\hat{m}$ the only non-vanishing term will be the $[\hat{H}_P,\hat{m}]$ commutator, which is proportional to $X(v)$. This 
means that if take the subset of the solutions where $X(v)\Psi=0$, the mass operator will be an observable. But if we look at 
the action of $X(v)$ in our new basis in \eref{pnewbasis} one will find that this is equivalent to the condition 
\eref{proccond1}. So $\hat{m}$ is an observable if the solutions are those which are equivalent to the solutions of the Proca 
field. But one may say that there are other solutions as well, since one does not have to impose \eref{proccond2}. The answer 
is that these states are special cases which are contained in the Proca solution. This is because in this case one has to 
solve $\sum_{I'}H_{I'I}(\underline{\Gamma})\tilde{C}_{I'}(\underline{\Gamma})$ for all $\Gamma$, which means that the solution 
will have to be in the kernel of all matrices appearing in $H_{I'I}$.\\
All in all the mass operator is an observable if if the solutions are those which are equivalent to the solutions of the 
Proca field. Since the Proca field did not have a potential term, the correspondence is correct only if we consider states 
where all $\Gamma$ are zero (note that other states the mass operator is also an observable, but it describes interactions).  

\section{Summary and open questions}

In this paper we analyzed the mass generation to a U(1) vector field via spontaneous symmetry breaking in LQG and 
compared the results obtained for the Proca field. Even at the classical level - after introducing new variables 
$\eta$ and $\Theta$ - the two theories had many similarities. The main difference was the extra scalar field and 
the potential term in the case of the spontaneous symmetry breaking, and where the Proca field had a mass parameter, 
we obtained a field. Thus it was not a surprise that the quantized theories were also similar, and in the case of 
spontaneous symmetry breaking mass became an operator. We defined a new basis in the quantum region, where the motivation 
was to find the eigenstates of the configuration variable of the scalar field. By choosing this new basis we were also 
able to rewrite the constraints to finite linear systems of differential equations, thus we were able to analyze only 
the scalar field dependence of the theory. We were able to (partially) solve the constraints and describe the behavior 
of the states when $\Gamma+a$ (i.e. the mass) tends to zero. We found that there exist states which are non-degenerate 
in this region, and further more at $\Gamma+a=0$ there is only one non-zero amplitude, the one which belongs to a 
solution to the ``gravity coupled to the electromagnetic field'' case.\\
The eigenvalues of the mass operator are continuous (though they have a discrete structure due to the discreteness 
appearing in the coefficient matrices) and real, but not necessarily positive (one needs extra input for this). 
A very interesting result is that the mass operator is an observable if the states are in the kernel of the 
corresponding momentum operator. This extra condition implies that in this case the other scalar field also becomes a gauge. 
This means that if we want a physically relevant mass operator, the scalar fields will not be real particles.\\
In the light of our results we can claim that - though the two theories are very similar - the spontaneous symmetry 
breaking has more advantages: 1. We are able to calculate the mass dependence of the states without solving the entire 
theory 2. We can produce the limit $m\to0$ without difficulty (the states have non-singular solutions) 3. In the case 
of the Proca field for different m we have different theories, while the spontaneous symmetry braking deals with all values. 
This is important if we want to calculate transition amplitudes between states that have different masses. 4. The mass 
is an eigenvalue of an operator which can be an observable, while in the case of the Proca field mass is a parameter.\\
In our analysis it was crucial that the scalar field had a commutative group, otherwise the eigenvalues of the configuration 
operator would have been hard to find. Thus it is an interesting question that in the case of non-commutative groups 
how can we generalize these results? But the commutative case also provides a few questions, like the complete analysis 
of the differential equation \eref{hamdiff}. The main question is what kind of restrictions do we have to make on the 
coefficient matrices to have a well defined, non-singular square-integrable solution.

\section{Acknowledgements}

This work was supported by the OTKA grant No. NI 68228. The author would also like to thank for the valuable help 
provided by Gyula Bene, Benjamin Beri, Janos Majar and Adam Rusznyak.

\section{Appendix A}

Here we introduce a notation which simplifies the scalar constraint. Let us introduce a multi-index I for the 
indices $s,f,\bar{\Gamma},\bar{\underline{\delta}}$ so that a type of expression 
$\langle s'|\langle f'|\langle\bar{\Gamma},\bar{\underline{\delta}}|\hat{X}|\bar{\Gamma},\bar{\underline{\delta}}\rangle |f\rangle |s\rangle$ 
will be denoted as $X_{I'I}$. Now consider those terms that do not contain $U_\eta$ or $\hat{X}_\eta$. 
These are $\hat{H}_{grav}$ and $\hat{H}_{YM}$, the Hamilton operator of the gravitational and Maxwell field. So in 
our new notation the contribution of these terms to the constraint equations will be the following: 

\begin{eqnarray}
\sum_{s',f',\underline{\lambda}',\underline{\bar{\delta}}'}\int d\Gamma(v)'\int d\bar{\Gamma}(v)'(H^G_{ss'}\delta_{ff'}\delta_{\underline{\lambda}\underline{\lambda}'}\delta_{\underline{\bar{\delta}}\underline{\bar{\delta}}'}\delta(\Gamma(v)',\Gamma(v))\delta(\bar{\Gamma}(v)',\bar{\Gamma}(v))+\nonumber\\
+H^{YM}_{ff'}G^1_{ss'}\delta_{\underline{\lambda}\underline{\lambda}'}\delta_{\underline{\bar{\delta}}\underline{\bar{\delta}}'}\delta(\Gamma(v)',\Gamma(v))\delta(\bar{\Gamma}(v)',\bar{\Gamma}(v)))C_{s',f',\underline{\lambda}',\underline{\bar{\delta}}'}(\underline{\Gamma}',\underline{\bar{\Gamma}}')=\nonumber\\
=\sum_{I'} H^{G+YM}_{I'I}(v)C_{I'}(\underline{\Gamma}),
\end{eqnarray}

where we performed the integration and sum on the variables related to the scalar fields.\\
The terms containing $U_\eta$ or $\hat{X}_\eta$ will be treated as follows: we shall write the dependence of these 
fields explicitly, while other expressions will be denoted (using the short notation) as $O^1_{I'I}$ etc. 
For example the contribution of the potential term $\hat{H}_{pot}$ will be denoted as follows:

\begin{eqnarray}
\sum_{s',f',\underline{\lambda}',\underline{\bar{\delta}}'}\int d\Gamma(v)'\int d\bar{\Gamma}(v)'\frac{1}{4}N(v)\mu \langle s|\hat{V}|s'\rangle\delta_{ff'}\delta(\bar{\Gamma},\bar{\Gamma}')\delta_{\bar{\underline{\delta}}\bar{\underline{\delta}}'}\delta_{\underline{\lambda}\underline{\lambda}'}\frac{1}{\lambda^2}\times\nonumber\\
\times C_{s',f',\underline{\lambda}',\underline{\bar{\delta}}'}(\underline{\Gamma}',\underline{\bar{\Gamma}}')\langle\Gamma,\underline{\lambda}|\arccos\Big(\frac{U_\eta(\lambda,v)+U^{-1}_\eta(\lambda,v)}{2}\Big)^2\times\nonumber\\
\times\Big[\frac{1}{\lambda}\arccos\Big(\frac{U_\eta(\lambda,v)+U^{-1}_\eta(\lambda,v)}{2}\Big)+2a\Big]^2|\Gamma',\underline{\lambda}'\rangle=\nonumber\\
=\sum_{I'}H^{pot}_{I'I}(v)C_{I'}(\underline{\Gamma})\Gamma(v)^2(\Gamma(v)+2a)^2
\end{eqnarray}

The derivative term $\hat{H}_{der}(\Theta)$ contains the operator $U_\eta$, so we have

\begin{eqnarray}
\sum_{s',f',\underline{\lambda}',\underline{\bar{\delta}}'}\int d\Gamma(v)'\int d\bar{\Gamma}(v)'\frac{1}{2}\frac{N(v)}{E(v)^2}\times\nonumber\\
\times \langle\Gamma,\underline{\lambda}|\Big[\frac{1}{\lambda}\arccos\Big(\frac{U_\eta (\lambda,v)+U^{-1}_\eta (\lambda,v)}{2}\Big)+a\Big]^{2}|\Gamma',\underline{\lambda}'\rangle\times\nonumber\\
\times\sum_{v(\Delta)=v(\Delta`)=v}\frac{1}{\delta^2}\langle\bar{\Gamma},\bar{\underline{\delta}}|[U_\Theta(\delta,s_n(\Delta))\underline{h}_{s_n}U_\Theta(\delta,v)^{-1}-1]\times\nonumber\\
\times[U_\Theta(\delta,s_r(\Delta'))\underline{h}_{s_r}U_\Theta(\delta,v)^{-1}-1]|\bar{\Gamma}',\bar{\underline{\delta}}'\rangle\langle s|\hat{G}_2^{nr}(v)|s'\rangle C_{s',f',\underline{\lambda}',\underline{\bar{\delta}}'}(\underline{\Gamma}',\underline{\bar{\Gamma}}')=\nonumber\\
=\sum_{I'}H^{der1}_{I'I}C_{I'}(\underline{\Gamma})(\Gamma(v)+a)^2
\end{eqnarray}

The other derivative term is a bit trickier since the $U_\eta$ is evaluated in different vertices. We have

\begin{eqnarray}
\sum_{s',f',\underline{\lambda}',\underline{\bar{\delta}}'}\int d\Gamma(v)'\int d\bar{\Gamma}(v)'\frac{1}{2}\frac{N(v)}{E(v)^2}\sum_{v(\Delta)=v(\Delta`)=v}\frac{1}{\lambda^2}\langle\Gamma,\underline{\lambda}|[U_\eta(\lambda,s_n(\Delta))U^{-1}_\eta(\lambda,v)-1]\times\nonumber\\
\times[U_\eta(\lambda,s_r(\Delta'))U^{-1}_\eta(\lambda,v)-1]|\Gamma',\underline{\lambda}'\rangle\langle s|\hat{G}_2^{nr}(v)|s'\rangle C_{s',f',\underline{\lambda}',\underline{\bar{\delta}}'}(\underline{\Gamma}',\underline{\bar{\Gamma}}')=\nonumber\\
=\sum_{I'}(H^{A}_{I'I}\exp(-2i\lambda\Gamma(v))-H^{B}_{I'I}\exp(-i\lambda\Gamma(v))+H^{C}_{I'I})C_{I'}(\underline{\Gamma}),
\end{eqnarray}

where the values $\exp(i\lambda\Gamma(s_r(\Delta')))$ etc. have been assimilated in the coefficients $H^{A}_{I'I}$ etc. 
since our differential equation will depend only variables in vertex v.\\
The last contribution is the momentum term. It is convenient to use equation \eref{eqgauge} so that one can simplify 
this expression in the following way:

\begin{eqnarray}
\sum_{s',f',\underline{\lambda}',\underline{\bar{\delta}}'}\int d\Gamma(v)'\int d\bar{\Gamma}(v)'\frac{1}{2}\frac{N(v)}{E(v)^2}(\langle\Gamma,\underline{\lambda}|X_\eta(v)^2|\Gamma',\underline{\lambda}'\rangle\delta_{\underline{\delta}\underline{\delta}'}\delta(\bar{\Gamma}(v)',\bar{\Gamma}(v))+\nonumber\\
+\langle\Gamma,\underline{\lambda}|\Big[\frac{1}{\lambda}\arccos\Big(\frac{U_\eta (\lambda,v)+U^{-1}_\eta (\lambda,v)}{2}\Big)+a\Big]^{-2})|\Gamma',\underline{\lambda}'\rangle\times\nonumber\\
\times \langle\bar{\Gamma},\bar{\underline{\delta}}|X_\Theta (v)^2)|\bar{\Gamma}',\bar{\underline{\delta}}'\rangle\langle s|\hat{G}_1(v)|s'\rangle\delta_{ff'}C_{s',f',\underline{\lambda}',\underline{\bar{\delta}}'}(\underline{\Gamma}',\underline{\bar{\Gamma}}')=\nonumber\\
\sum_{s',f',\underline{\lambda}',\underline{\bar{\delta}}'}\int d\Gamma(v)'\int d\bar{\Gamma}(v)'\frac{1}{2}\frac{N(v)}{E(v)^2}(\langle\Gamma,\underline{\lambda}|X_\eta(v)^2|\Gamma',\underline{\lambda}'\rangle+\nonumber\\
+\frac{L(v)^2}{(\Gamma(v)+a)^2}\delta(\Gamma(v)',\Gamma(v))\delta_{\underline{\lambda}\underline{\lambda}'})\langle s|\hat{G}_1(v)|s'\rangle\delta_{\underline{\delta}\underline{\delta}'}\delta(\bar{\Gamma}(v)',\bar{\Gamma}(v))\delta_{ff'}C_{s',f',\underline{\lambda}',\underline{\bar{\delta}}'}(\underline{\Gamma}',\underline{\bar{\Gamma}}')=\nonumber\\
=\sum_{I'}((\lambda_v-i\frac{\delta}{\delta\Gamma(v)})^2+\frac{L(v)^2}{(\Gamma(v)+a)^2})H^P_{I'I}C_{I'}(\underline{\Gamma}),
\end{eqnarray}

where $L(v)=\sum_v l_e$ is the sum 

\section{Appendix B}

Here we describe the method to solve a system of differential equation of the form
\begin{eqnarray}
\ddot{\vec{c}}(t)={\bf H}(t)\vec{c}(t),
\end{eqnarray}

where ${\bf H}(t)$ is a $N\times N$ matrix.\\ 
The method is similar to the one used in cases of path ordered integration. First we integrate the equation:
\begin{eqnarray}
\dot{\vec{c}}(t)=\int_0^t dt_1{\bf H}(t_1)\vec{c}(t_1)+\vec{c}_1,
\end{eqnarray}

where $$\vec{c}_1=\dot{\vec{c}}(0).$$ Another integration yields

\begin{eqnarray}
\vec{c}(t)=\int_0^t dt_1\int_0^{t_1} dt_2{\bf H}(t_2)\vec{c}(t_2)+\vec{c}_1 t+\vec{c}_0,
\end{eqnarray}

where $$\vec{c}_0=\vec{c}(0).$$ Now we iterate this equation and arrive to the result

\begin{eqnarray}
\vec{c}(t)=\Big(1+\sum_{j=1}^\infty\int_0^t dt_1\int_0^{t_1} dt_2\dots\int_0^{t_{2j-1}}dt_{2j}{\bf H}(t_2){\bf H}(t_4)\dots {\bf H}(t_{2j})\Big)\vec{c}_0+\nonumber\\
+\Big(t+\sum_{j=1}^\infty\int_0^t dt_1\int_0^{t_1} dt_2\dots\int_0^{t_{2j-1}}dt_{2j}{\bf H}(t_2){\bf H}(t_4)\dots {\bf H}(t_{2j})t_{2j}\Big)\vec{c}_1
\end{eqnarray}

\end{document}